\documentclass[prl,twocolumn,showpacs,amsmath,amssymb,superscriptaddress]{revtex4}

\usepackage{graphicx}    
\usepackage{dcolumn}     
\usepackage{bm}          
\usepackage{subfigure}   
\usepackage[normalem]{ulem}
\usepackage{xcolor}
\newcommand{\hide} [1] { }                                      
\graphicspath{{Figures/},.}
\newcommand{\ethz}{ETH Zurich, CLU E1, Clasiusstrasse 50, 8092 Zurich, Switzerland}
\begin{document}


\title{On Phase Transitions to Cooperation in the Prisoner's Dilemma}

\author{Dirk Helbing}
\affiliation{\ethz}
\author{Sergi Lozano}
\affiliation{\ethz}
\begin{abstract}
Game theory formalizes certain interactions between physical particles or between living beings in biology, sociology, and economics, and quantifies the outcomes by payoffs. The prisoner's dilemma (PD) describes situations in which it is profitable if everybody cooperates rather than defects (free-rides or cheats), but as cooperation is risky and defection is tempting, the expected outcome is defection.  Nevertheless, some biological and social mechanisms can support cooperation by effectively transforming the payoffs. Here, we study the related phase transitions, which can be of first order (discontinous) or of second order (continuous), implying a variety of different routes to cooperation. After classifying the transitions into cases of equilibrium displacement, equilibrium selection, and equilibrium creation, we show that a transition to cooperation may take place even if the stationary states and the eigenvalues of the replicator equation for the PD stay unchanged.
Our example is based on adaptive group pressure, which makes the payoffs dependent on the endogeneous dynamics in the population. The resulting bistability can invert the expected outcome in favor of cooperation.
\end{abstract}

\date{\today}

\maketitle

{\it Introduction.} 
Game theory goes back to von Neumann \cite{Neumann}, one of the superminds of quantum mechanics. Originally intended to describe interactions in economics, sociology, and biology \cite{Neumann,Axelrod,gamedyn}, it has recently become a quickly growing research area in physics, where methods from non-linear dynamics and pattern formation \cite{segr,PNAS}, agent-based or particle-like models \cite{PNAS,EPL}, network theory \cite{Network2} and statistical physics \cite{Engel} are applied. There are even quantum theoretical contributions \cite{quantum1}. 
\par
When two entities characterized by the states, ``strategies'', or ``behaviors'' $i$ and $j$ interact with each other,  game theory formalizes the result by payoffs $P_{ij}$, and the structure of the payoff matrix $(P_{ij})$ determines the kind of the game. The dynamics of a system of such entities is often delineated by the so-called replicator equations 
\begin{equation}
\frac{d p(i,t)}{dt} = p(i,t) \bigg[\!  \sum_j P_{ij}p(j,t) - \! \sum_{j,l} p(l,t) P_{lj} p(j,t) \bigg]  
\label{eq:repl}
\end{equation}
\cite{gamedyn}. $p(i,t)$ represents the relative frequency of behavior $i$ in the system, which increases when the expected ``success'' $F_i=\sum_j P_{ij}p(j,t)$ exceeds the average one, $\sum_i F_i p(i,t)$. 
\par
Many collective phenomena in physics such as agglomeration or segregation phenomena can be studied in a game-theoretical way \cite{PNAS,EPL}. Applications also include the theory of evolution \cite{Eigen} and the study of ecosystems \cite{Ecosystems}. Another exciting research field is the study of mechanisms supporting the cooperation between selfish individuals \cite{Neumann,Axelrod,gamedyn} in situations like the ``prisoner's dilemma'' or public goods game, where they would usually defect (free-ride or cheat). Contributing to public goods and sharing them constitute ubiquitous situations, where cooperation is crucial, for example, in order to maintain a sustainable use of natural resources or a well-functioning health or social security system.
\par 
In the following, we will give an overview of the stationary solutions of the replicator equations (\ref{eq:repl}) and their stability properties. Based on this, we will discuss several ``routes to cooperation'', which transform the prisoner's dilemma into other games via different sequences of continuous or  discontinuous phase transitions. These routes will then be connected to different biological or social mechanisms accomplishing such phase transitions \cite{Five}.
Finally, we will introduce the concept of  ``equilibrium creation'' and distinguish it from routes to cooperation based on ``equilibrium selection'' or ``equilibrium displacement''. A new cooperation-promoting mechanism based on adaptive group pressure will exemplify it.
\par
{\it Stability properties of different games.} 
Studying games with two strategies $i$ only, the replicator equations (\ref{eq:repl}) simplify, and we remain with 
\begin{equation}
\frac{dp(t)}{dt}  = p(t) [1-p(t)] \big\{ \lambda_1 [1-p(t)] - \lambda_2 \, p(t) \big\} \, , 
\label{rep3}
\end{equation}
where $p(t) = p(1,t)$ represents the fraction of cooperators and $1-p(t)=p(2,t)$ the fraction of defectors.
$\lambda_1 = P_{12} -P_{22}$ 
and $\lambda_2 = P_{21} - P_{11}$ 
are the eigenvalues of the two stationary solutions $p=p_1 = 0$ and $p=p_2=1$. If $0< \lambda_1 / (\lambda_1 + \lambda_2) < 1$, there is a third stationary solution $p=p_3 = \lambda_1 / (\lambda_1+\lambda_2)$ with eigenvalue 
$\lambda_3  = -(1-p_3)\lambda_1$. For the sake of our discussion, we imagine an additional fluctuation term $\xi(t)$ on the right-hand-side of Eq. (\ref{rep3}), reflecting small perturbations of the strategy distribution. 
\par
Four different cases can be classified \cite{gamedyn}: (1) If $\lambda_1<0$ and $\lambda_2 >0$, the stationary solution $p_1$ corresponding to defection by everybody is stable, while the stationary solution $p_2$ corresponding to cooperation by everyone is unstable. That is, any small perturbation will drive the system away from full cooperation towards full defection. This situation applies to the {\it prisoner's dilemma (PD)} defined by payoffs with $P_{21} > P_{11} > P_{22} > P_{12}$. According to this, strategy $i=1$ (``cooperation'') is risky, as it can yield the lowest payoff $P_{12}$, while strategy $i=2$ (``defection'') is tempting, since it can give the highest payoff $P_{21}$.
(2) If $\lambda_1>0$ and $\lambda_2 <0$, the stationary solution $p_1$ is unstable, while $p_2$ is stable. This means that the system will end up with cooperation by everybody. Such a situation occurs for the so-called {\it harmony game (HG)} with  $P_{11} > P_{21} > P_{12} > P_{22}$, as mutual cooperation gives the highest payoff $P_{11}$. (3) If $\lambda_1>0$ and $\lambda_2 >0$,
the stationary solutions $p_1$ and $p_2$ are unstable, but there exists a third stationary solution $p_3$, which turns out to be stable. As a consequence, the system is driven towards a situation, where a fraction $p_3$ of cooperators is expected to coexist with a fraction $(1-p_3)$ of defectors. Such a situation occurs for the {\it snowdrift game (SD)} (also known as hawk-dove or chicken game). This game is characterized by $P_{21} > P_{11} > P_{12} > P_{22}$ and assumes that unilateral defection is tempting, as it yields the highest payoff $P_{21}$, but also risky, as mutual defection gives the lowest payoff $P_{22}$. 
(4) If $\lambda_1<0$ and $\lambda_2<0$, the stationary solutions $p_1$ and $p_2$ are both stable, while the stationary solution $p_3$ is unstable. As a consequence, full cooperation is possible, but not guaranteed. In fact, the final state of the system depends on the initial condition $p(0)$ (the ``history''): If $p(0) < p_3$,  the system is expected to end up in the stationary solution $p_1$, i.e. with full defection. If $p(0)>p_3$, the system is expected to move towards $p_2=1$, corresponding to cooperation by everybody. The history-dependence implies that the system is multistable (here: bistable), as it has several (locally) stable solutions. This case is found for the {\it stag hunt game (SH)} (also called assurance). This game is characterized by $P_{11} > P_{21} > P_{22} > P_{12}$, i.e. cooperation is rewarding, as it gives the highest payoff $P_{11}$ in case of mutual cooperation, but it is also risky, as it yields the lowest payoff $P_{12}$, if the interaction partner is uncooperative.
\par
{\it Phase transitions and routes to cooperation.} 
When facing a prisoner's dilemma, it is of vital interest to transform the payoffs in such a way that cooperation between individuals is supported. Starting with the payoffs $P_{ij}^0$ of a prisoner's dilemma, one can reach different payoffs $P_{ij}$, for example, by introducing strategy-dependent taxes $T_{ij} = P_{ij}^0-P_{ij}>0$. When increasing the taxes $T_{ij}$ from 0 to $T_{ij}^0$, the eigenvalues will change from $\lambda_1^0= P_{12}^0 -P_{22}^0$ and $\lambda_2^0= P_{21}^0 - P_{11}^0$ to $\lambda_1 =  \lambda_1^0 +T_{22} -T_{12}$ 
and $\lambda_2 = \lambda_2^0 +T_{11}-T_{21}$. In this way, one can 
create a variety of routes to cooperation, which are characterized by different kinds of phase transitions. We define {\it route 1} [PD$\rightarrow$HG] by a direct transition from a prisoner's dilemma to a harmony game. It is characterized by a discontinuous transition from a system, in which defection by everybody is stable, to a system, in which cooperation by everybody is stable (see Fig. 1a). {\it Route 2} [PD$\rightarrow$SH] is defined by a direct transition from the prisoner's dilemma to a stag hunt game. After the moment $t_*$, where $\lambda_2$ changes from positive to negative values, the system behavior becomes history-dependent: When the fluctuations $\xi(t)$ for $t>t_*$ exceed the critical threshold $p_3(t) = \lambda_1/[\lambda_1+\lambda_2(t)]$, the system will experience a sudden transition to cooperation by everybody. Otherwise one will find defection by everyone, as in the prisoner's dilemma (see Fig. 1b). In order to make sure that the perturbations $\xi(t)$ will eventually exceed $p_3(t)$ and  trigger cooperation, the value of $\lambda_2$ must be reduced to sufficiently large negative values. It is also possible to have a continuous rather than sudden transition to cooperation: We define {\it route 3} [PD$\rightarrow$SD] by a transition from a prisoner's dilemma to a snowdrift game. As $\lambda_1$ is changed from negative to positive values, a fraction $p_3(t) = \lambda_1(t)/[\lambda_1(t)+\lambda_2]$ of cooperators is expected to result (see Fig. 1c). When increasing $\lambda_1$, this fraction rises continuously. One may also implement more complicated transitions. {\it Route 4}, for example, establishes the transition sequence PD$\rightarrow$SD$\rightarrow$HG (see Fig. 1d), while we define {\it route 5} by the transition PD$\rightarrow$SH$\rightarrow$HG (see Fig. 1e). One may also implement the transition PD$\rightarrow$SD$\rightarrow$HG$\rightarrow$SH ({\it route 6}, see Fig. 1f), establishing a path-dependence, which can guarantee cooperation by everybody in the end. (When using {\it route 2}, the system remains in a defective state, if the perturbations do not exceed the critical value $p_3$.) 
\par\begin{figure}[htp]
\begin{center}
\includegraphics[width=8.7cm]{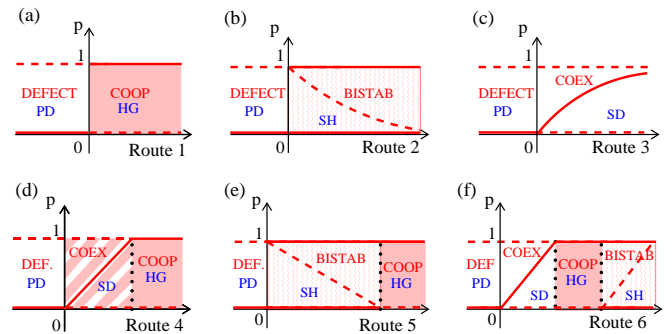}
\end{center}
\caption[]{Schematic illustration of the phase transitions defining the different routes to cooperation. The order parameter is the stationary frequency of cooperators, while the control parameters are the parameters $r$, $w$, $k$, $m$, or $q$ in Nowak?s cooperation-enhancing rules \cite{Five} (see main text) or, more generally, (non-)linear combination of the model parameters $b$ and $c$. Solid (red) lines represent {\it stable} stationary proportions of cooperators, dashed lines {\it unstable} fix points. Diagonal lines show the additional stationary solution $p_3$, where $0\le p_3 \le 1$. ($p$ = proportion of cooperators; DEFECT = defection is stable, i.e. everybody defects; COOP = cooperation is stable, i.e. everybody cooperates; COEX = mixture of defectors with a proportion $p_3$ of cooperators; BISTAB = cooperation is stable if $p_3 < p(0)$, where $p(0)$ means the initial proportion of cooperators, otherwise everybody defects.)} 
\label{fig1}
\end{figure}

{\it Relationship with cooperation-supporting mechanisms.}
We will now discuss the relationship of the above introduced routes to cooperation with biological and social mechanisms (``rules'') promoting the evolution of cooperation. Martin A. Nowak performs his analysis of five such rules with the reasonable specifications $T=b>0$, $R=b-c>0$, $S=-c<0$, and $P=0$ in the limit of weak selection \cite{Five}. Cooperation is assumed to require a contribution $c>0$ and to produce a benefit $b>c$ for the interaction partner, while defection generates no payoff ($P=0$).  As most mechanisms leave $\lambda_1$ or $\lambda = (\lambda_1+\lambda_2)/2$ unchanged, we will now focus on the payoff-dependent parameters $\lambda_1$ and $\lambda$ (rather than $\lambda_1$ and $\lambda_2$). The basic prisoner's dilemma is characterized by $\lambda_1^0 = -c$ and $\lambda^0 = 0$. 
\par
According to the Supporting Online Material of Ref. \cite{Five}, 
{\it kin selection} (genetic relatedness) tranforms the payoffs into 
$P_{11} = P^0_{11} + r (b-c)$, $P_{12} = P^0_{12}+br$, $P_{21} = P^0_{21}-cr$, and $P_{22} = P^0_{22}$.
Therefore, it leaves $\lambda$ unchanged and increases $\lambda_1$
by $T_{22}-T_{12} = br$, where $r$ represents the degree of genetic relatedness.
{\it Direct reciprocity} (repeated interaction) does not change $\lambda_1$, but it reduces
$\lambda$ by $-\frac{1}{2}(b-c)[1/(1-w)-1]<0$, where $w$ is the
probability of a future interaction. {\it Network reciprocity}
(clustering of individuals playing the same strategy) leaves
$\lambda$ unchanged and increases $\lambda_1$ by $H(k)$, where
$H(k)$ is a function of the number $k$ of neighbors. Finally, {\it
group selection} (competition between different populations)
increases $\lambda_1$ by $(b-c)(m-1)$, where $m$ is the number of
groups, while $\lambda$ is not modified. However, $\lambda_1$ and $\lambda$ may also change simultaneously. For example, {\it indirect reciprocity} (based on trust and reputation) increases $\lambda_1$ by $cq$ and reduces $\lambda$ by $-\frac{1}{2}(b-c)q<0$, where $q$ quantifies social acquaintanceship. 
\par
Summarizing this, kin selection, network reciprocity, and
group selection preserve $\lambda = 0$ and increase the value of
$\lambda_1$ (see route 1 in Fig. 2). Direct reciprocity, in contrast, preserves the value of $\lambda_1$ and reduces $\lambda$ (see route 2a in Fig. 2). Indirect reciprocity promotes the same transition (see route 2b in Fig. 2). 
Supplementary, one can analyze {\it costly punishment}. Using the payoff specifications made in the Supporting Information of Ref. \cite{Punish}, costly punishment changes $\lambda$ by $-(\beta+\gamma)/2 <0$ and $\lambda_1$ by $-\gamma$ \cite{Punish}, i.e. when $\gamma$ is increased, the values of $\lambda$ and $\lambda_1$ are simultaneously reduced (see route 2c in Fig. 2). Here, $\gamma>0$ represents the punishment cost invested by a cooperator to impose a punishmet fine $\beta > 0$ on a defector, which decreases the payoffs of both interaction partners.
Route 3 can be generated by the formation of friendship networks \cite{JTB}. Route 4 may occur by kin selection, network reciprocity, or group selection, when starting with a prisoner's dilemma with $\lambda^0 <0$ (rather than $\lambda^0 =0$ as assumed before). Route 5 may be generated by the same mechanisms, if $\lambda^0 > 0$. Finally, route 6 can be implemented by time-dependent taxation (see Fig. 2). 
\par\begin{figure}[htp]
\begin{center}
\includegraphics[width=8.4cm]{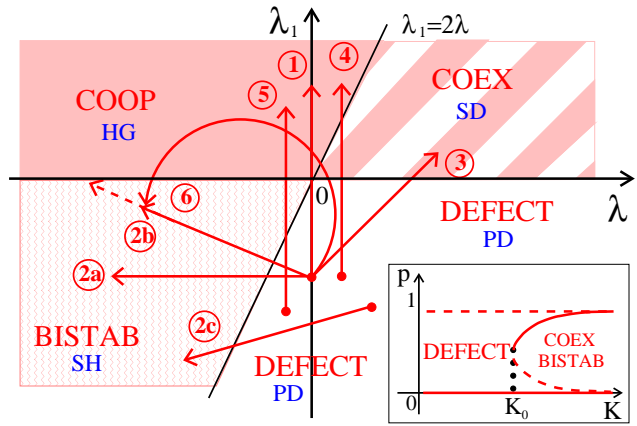}
\end{center}
\caption[]{Phase diagram of expected system behaviors, based on an analysis of the game-dynamical replicator equation (2) as a function of the parameters $\lambda$ and $\lambda_1$. The different routes to cooperation are illustrated by arrows.  Terms in capital letters are defined in Fig. 1. Inset: Stable stationary solutions (solid lines) and unstable ones (broken lines) as functions of the parameter $K$, when the reward depends on the proportion of cooperators. The bifurcation at the ``tipping point'' $K=K_0$ ``inverts'' the system behavior (see main text).}
\label{fig2}
\end{figure}
\par
{\it Further kinds of transitions to cooperation.}
The routes to cooperation discussed so far change the eigenvalues $\lambda_1$ and $\lambda_2$, and leave the stationary solutions $p_1$ and $p_2$ unchanged. However, transitions to cooperation can also be generated by shifting the stationary solutions or creating new ones, as we will show now. For this, we generalize the replicator equation (\ref{rep3}) by replacing $\lambda_1$ with $f(p)$ and $\lambda$ with $g(p)$, and by adding a term $h(p)$, which can describe effects of spontaneous transitions like mutations. To guarantee $0 \le
p(t) \le 1$, we must have $h(p) = v(p) - p w(p)$ with functions $w(p)\ge v(p) \ge 0$. The
resulting equation is $dp/dt = F(p(t))$ with $F\big(p\big) = (1-p) [ f(p) - 2 g(p) p ] p + h(p)$, 
and its stationary solutions $p_k$ are given by $F(p_k) = (1-p_k) [ f(p_k) - 2g(p_k) p_k] p_k + h(p_k) = 0$. The associated eigenvalues $\lambda_k = dF(p_k)/dp$ determining the stability of the stationary solutions $p_k$ are 
\begin{equation}
\lambda_k = (1-2p_k) ( f_k - 2p_kg_k )  + p_k (1-p_k)( f'_k - 2p_k g'_k  - 2 g_k ) + h'_k, \nonumber
\end{equation}
where we have used the abbreviations $f_k = f(p_k)$, $g_k = g(p_k)$, $h_k = h(p_k)$. $f'_k = f'(p_k)$, $g'_k = g'(p_k)$ and $h_k = h'(p_k)$ are the derivatives of the functions $f(p)$, $g(p)$ and $h(p)$ in the points $p=p_k$.
\par
{\it Classification.} We can now distinguish different {\it kinds} of transitions from defection to cooperation: If the stationary solutions $p_1 = 0$ and $p_2=1$ of the prisoner's dilemma are modified, we talk about transitions to cooperation by {\it equilibrium displacement}. This case occurs, for example, when random mutations are not weak ($h\ne 0$). If the eigenvalues $\lambda_1$ or $\lambda_2$ of the stationary solutions $p_1=0$ and $p_2=1$ are changed, we speak of {\it equilibrium selection.} This case applies to all routes to cooperation discussed before. If a new stationary solution appears, we speak of {\it equilibrium creation}. 
The different cases often appear in combination with each other (see the Summary below). In the following, we will discuss an interesting case, where cooperation occurs solely through equilibrium creation, i.e. the stationary solutions $p_1$ and $p_2$ of the replicator equation for the prisoner's dilemma as well as their eigenvalues $\lambda_1$ and $\lambda_2$ remain unchanged. 
We illustrate this by the example of an {\it adaptive kind of group pressure} that rewards mutual cooperation ($T_{11} < 0$) or sanctions unilateral defection ($T_{21} > 0$). Both, rewarding and sanctioning reduces the value of $\lambda_2$, while $\lambda_1$ remains unchanged. Assuming here that the group pressure vanishes, when everybody cooperates (as it is not needed then), while it is maximum when everybody defects (to encourage cooperation) \cite{Press}, we may set $f(p)=\lambda_1^0$ and $g(p) = \lambda^0 - K[1-p(t)]$, corresponding to $\lambda_2(t) = \lambda_2^0 - 2K [1-p(t)]$.  
It is obvious that we still have the two stationary solutions $p_1=0$ and $p_2=1$ with the eigenvalues $\lambda_1 = \lambda_1^0<0$ and $\lambda_2 = 2\lambda^0-\lambda_1^0>0$ of the original prisoners dilemma with parameters $\lambda_1^0$ and $\lambda_2^0$ or $\lambda^0$. However, for large enough values of $K$ [namely for $K> K_0 = \lambda^0 + |\lambda_1^0|+ \sqrt{|\lambda_1^0|(2\lambda^0 + |\lambda_1^0|)}$], we find two additional stationary solutions
\begin{equation}\textstyle
p_\pm = \frac{1}{2} - \frac{\lambda^0}{2K} \pm \sqrt{\left( \frac{1}{2} - \frac{\lambda^0}{2K} \right)^2 - \frac{|\lambda_1^0|}{2K} } \, .
\label{newsol}
\end{equation}
$p_-$ is an {\it unstable} stationary solution with $p_1<p_- < p_+$ and $\lambda_- = dF(p_-)/dp > 0$, while $p_+$ is a {\it stable} stationary solution with $p_-<p_+ < p_2$ and $\lambda_+ = dF(p_+)/dp < 0$ (see inset of Fig. 2). Hence, the assumed dependence of the payoffs on the proportion $p$ of cooperators generates a {\it bistable} situation (BISTAB), with the possibility of a coexistence of a few defectors with a large proportion $p_+$ of cooperators, given $K>K_0$. If $p(0)< p_-$, where $p(0)$ denotes the initial condition, defection by everybody results, while a stationary proportion $p_+$ of cooperators is established for $p_-<p(0) < 1$. Surprisingly, in the limit $K\rightarrow \infty$, cooperation is established for {\it any} initial condition $p(0) \ne 0$ (or through fluctuations).
\par
{\it Summary.} 
We have discussed from a physical point of view what must happen that social or biological, payoff-changing interaction mechanisms can create cooperation in the prisoner's dilemma. The possible ways are (i) moving the stable stationary solution away from pure defection (routes 3, 4, and 6), (ii) stabilizing the unstable solution (routes 1, 2, 4, 5 and 6), or (iii) creating new stationary solutions, which are stable (routes 3, 4 and 6). Several of these points can be combined. If (i) is fulfilled, we speak of  ``equilibrium displacement'', if their eigenvalues change, we called this  ``equilbrium selection'', and if (iii) is the case, we talk of ``equilibrium creation''. The first case can result from mutations, the second one applies to many social or biological cooperation-enhancing mechanisms \cite{Five}. We have discussed an interesting case of equilibrium creation, in which the outcome of the replicator equation is changed, although the stationary solutions of the PD and their eigenvalues remain unchanged. This can, for example, occur by adaptive group pressure \cite{Press}, which introduces an adaptive feedback mechanism and thereby increases the order of non-linearity of the replicator equation. Surprisingly, already a linear dependence of the payoff values $P_{ij}$ on the endogeneous dynamics $p(t)$ of the system is enough to destabilize defection and stabilize cooperation, thereby inverting the outcome of the prisoner's dilemma.
\par

{\it Acknowledgments.}
This work was partially supported by the Future and Emerging Technologies programme FP7-COSI-ICT of the European Commission through the project QLectives (grant no.: 231200).

\end{document}